\documentclass[twocolumn,showpacs,amsmath,amssymb]{revtex4}
\usepackage{graphicx}
\usepackage{dcolumn}
\usepackage{bm}
\newcommand{\beq}{\begin{equation}}
\newcommand{\eeq}{\end{equation}}
\newcommand{\bea}{\begin{eqnarray}}
\newcommand{\eea}{\end{eqnarray}}
\newcommand{\eps}{\varepsilon}

\begin{document}

\author{E. E. Saperstein}
\affiliation{Kurchatov Institute, 123182 Moscow, Russia}
\affiliation{National Research Nuclear University MEPhI, 115409
Moscow, Russia}

\author{M. Baldo}
\affiliation{INFN, Sezione di Catania, 64 Via S.-Sofia, I-95125 Catania, Italy}

\author{S. S. Pankratov}
\affiliation{Kurchatov Institute, 123182 Moscow, Russia}
\affiliation{Moscow Institute of Physics and Technology, 141700
Dolgoprudny, Russia.}

\author{S. V. Tolokonnikov}
\affiliation{Kurchatov Institute, 123182 Moscow, Russia}
\affiliation{Moscow Institute of Physics and Technology, 141700
Dolgoprudny, Russia.}

\title{Including particle-vibration coupling in the Fayans functional.
Odd-even mass differences of semi-magic nuclei.}


\begin{abstract}A method to evaluate the particle-phonon coupling (PC) corrections to the single-particle
energies in  semi-magic nuclei, based on the direct solution of the Dyson equation with PC corrected
mass operator, is presented. It is used for finding the odd-even mass difference between even Pb and
Sn isotopes and their odd-proton neighbors.  The Fayans energy density functional (EDF) DF3-a is used
which gives rather highly accurate predictions for these mass differences already at the mean-field
level. In the case of the lead chain, account for the PC corrections induced by the low-laying phonons
$2^+_1$ and $3^-_1$ makes agreement of the theory with the experimental data significantly better. For
the tin chain, the situation is not so definite. In this case,  the PC corrections make agreement
better in the case of the addition mode but they spoil the agreement for the removal mode.  We discuss
the reason of such a discrepancy.   \end{abstract}

\maketitle

\section{Introduction}
 Ability to describe nuclear single-particle (SP) spectra correctly is an important feature of any
self-consistent nuclear theory. Last decade is characterized with a revival of the interest to study
the particle-phonon coupling (PC) corrections to the SP spectra within different self-consistent
nuclear approaches. Up to now, they were limited only to magic nuclei, see a study of this problem
within  the relativistic mean-field theory \cite{Litv-Ring}, and several ones within the
Skyrme--Hartree--Fock method \cite{Bort,Dobaczewski,Baldo-PC}.  The consideration of this problem
\cite{levels} on the basis of the self-consistent theory of finite Fermi systems (TFFS) \cite{scTFFS}
differs from those cited above by the account for the non-pole terms of the PC correction $ \delta
\Sigma^{\rm PC}$ to the mass operator $\Sigma_0$.  In the nuclear PC problem, the non-pole diagrams of
the operator $ \delta \Sigma^{\rm PC}$  were considered firstly by Khodel \cite{Khod-76}, see also
\cite{scTFFS}.  These diagrams  are also sometimes named ``the phonon tadpole'' \cite{phon-tad}, by
the analogy with the  tadpole-like   diagrams in field theory \cite{tad}.

 There are several reasons, why the magic nuclei were chosen as a polygon for studying  the PC corrections
to SP levels. First, the experimental SP levels are known in detail only for magic nuclei
\cite{lev-exp}. Second, these nuclei are non-superfluid, which makes the formulas for the PC
correction $\delta \Sigma^{\rm PC}$ to the mass operator much simpler than their analogue in the
superfluid case \cite{phon-tad}. At last, in these nuclei, the PC strength is rather week and the
perturbation theory in $g_L^2$ is valid, $g_L$ being the creation vertex of the $L$-phonon. Moreover,
the perturbation theory in $\delta \Sigma^{\rm PC}$ can be used  for solving the Dyson equation with
the mass operator $\Sigma^{\rm PC}(\eps){=}\Sigma_0{+}\delta \Sigma^{\rm PC}(\eps)$.

Another situation occurs often in  semi-magic nuclei \cite{lev-semi}, due to a strong mixture of some
SP states $|\lambda\rangle$ with those possessing the structure of ($|\lambda\rangle$ + $L$-phonon).
For such cases a method was   used  in \cite{lev-semi} based on the direct solution of the Dyson
equation with the mass operator $\Sigma^{\rm PC}(\eps)$.  Such a method was developed by Ring and
Werner in 1973 \cite{Ring-Werner}. As result of such solution each SP state $|\lambda\rangle$ splits
to a set of $|\lambda,i\rangle$ solutions with the SP strength distribution factors $S^i_{\lambda}$.
An approximate expression for the non-pole term of $\delta \Sigma^{\rm PC}$ was used in
\cite{lev-semi}, just as in  \cite{levels}, which is valid for heavy nuclei.

A method was proposed, how to express the average SP energy $\eps_{\lambda}$ and the average
$Z_{\lambda}$ factor in terms of  $\eps_{\lambda}^i$ and $S_{\lambda}^i$. It is similar to the one
used usually for finding the corresponding experimental values \cite{lev-exp}. Unfortunately, the
experimental data in heavy non-magic nuclei considered in \cite{lev-semi} are practically absent.
However, there is a massive set of data which has a direct relevance to the solutions
$\eps_{\lambda}^i$ under discussion, namely, the odd-even mass differences, that is the ``chemical
potentials'' in  terms  of the TFFS \cite{AB}. This quantity was analyzed in  in a brief letter
\cite{chim-pots} for the lead chain.  In this article, we present a detailed analysis of these results
and carry out the analogous calculations for the tin chain.

\section{Direct solution of the Dyson equation with PC corrected mass operator}

In this article, just as in Refs. \cite{levels,lev-semi,chim-pots}, we use the self-consistent basis
generated with the energy density functional (EDF) by Fayans \cite{Fay1,Fay4,Fay}: \beq E_0=\int d^3 r
{\cal E}\left(\rho ({\bf r}),\nu ({\bf r)}\right),  \label{E0}\eeq which depends on the equal footing
on the normal density $\rho$ and the anomalous one $\nu$. Note, that from the very beginning Fayans
with the coauthors supposed that the parameters of the EDF should be chosen in such a way that they
take into account various PC corrections on average.

To find the SP energies with account for the PC effects, we solve the following equation: \beq \left(\eps-H_0
-\delta \Sigma^{\rm PC}(\eps) \right) \phi =0, \label{sp-eq}\eeq where $H_0$ is the quasiparticle
Hamiltonian with the spectrum $\eps_{\lambda}^{(0)}$ and $\delta \Sigma^{\rm PC}$ is the PC correction
to the quasiparticle mass operator. This is equivalent to the Dyson equation for the one-particle
Green function $G$ with the mass operator $\Sigma(\eps)=\Sigma_0+\delta \Sigma^{\rm PC}(\eps)$.

For brevity, we present below explicit formulas for one $L$-phonon.
In the case when several $L$-phonons are taken into account, the total PC variation of the mass
operator in Eq. (\ref{sp-eq}) is  the sum over all phonons: \beq \delta \Sigma^{\rm PC} = \sum_L
\delta \Sigma^{\rm PC}_L . \label{sum-L}\eeq

The corresponding diagrams
for the PC correction $\delta \Sigma^{\rm PC}_L$ to the mass operator $\Sigma_0$, in the
representation of the SP states $|\lambda\rangle$, are displayed in Fig. 1. The first one is the usual
pole diagram, with obvious notation, whereas the second one represents the sum of all non-pole
diagrams of the order $g_L^2$.

\begin{figure}
\centerline {\includegraphics [width=80mm]{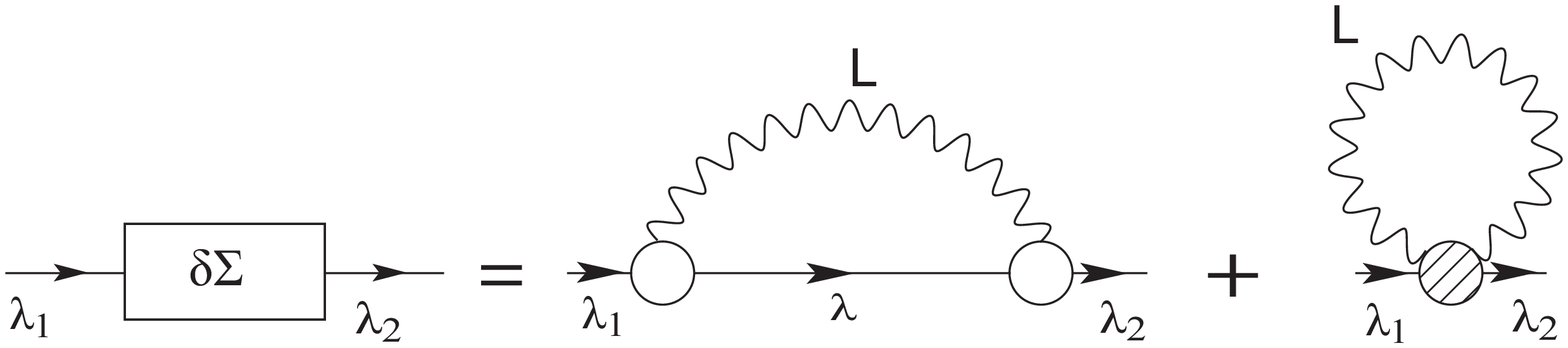}} \caption{PC corrections to the mass
operator. The open circle is the vertex $g_L$ of the $L$-phonon creation. The gray blob denotes
the phonon non-pole (``tadpole'') term.} \label{fig:SigPC}
\end{figure}

In magic nuclei, the vertex $g_L$  in Fig. 1  obeys the RPA-like TFFS equation
\cite{AB} \beq { g_L}(\omega)={{\cal F}} {A}(\omega) {
g_L}(\omega), \label{g_L} \eeq
 where $ A(\omega)=\int
G \left(\eps + \omega/ 2 \right) G \left(\eps - \omega/ 2 \right)d
\eps/(2 \pi i)$ is the particle-hole propagator and ${\cal F}$ is the Landau-Migdal (LM) interaction
amplitude.

The expression for the pole term in magic nuclei is well known \cite{Litv-Ring,Bort,levels}, but we
present it here for completeness. In obvious symbolic notation, the pole diagram corresponds to
$\delta\Sigma^{\rm pole}_L=(g_L,D_LGg_L)$, where $D_L(\omega)$ is the $L$-phonon $D$-function: \beq
D_L(\omega)= \frac 1 {\omega - \omega_L +i\gamma} - \frac 1 {\omega + \omega_L - i\gamma}
\,,\label{DL} \eeq where $\omega_L$ is the $L$-phonon excitation energy.

Explicitly, one obtains
\bea \delta\Sigma^{\rm
pole}_{\lambda_1\lambda_2}(\epsilon)&=&\sum_{\lambda\,M}
\langle\lambda_1|g_{LM}|\lambda\rangle  \langle\lambda|g^+_{LM}|\lambda_2\rangle \nonumber\\
&\times&\left(\frac{n_{\lambda}}{\eps+\omega_L-
\eps_{\lambda}^{(0)}}+\frac{1-n_{\lambda}}{\eps-\omega_L
-\eps_{\lambda}^{(0)}}\right), \label{dSig-pole} \eea where $n_{\lambda}=(0,1)$
stands for the occupation numbers.

As to the non-pole term,
we follow to the method developed  by Khodel \cite{Khod-76}, see also \cite{scTFFS}.
The second, non-pole, term in Fig. 1   is \beq \delta\Sigma^{\rm
non-pole}=\int \frac {d\omega} {2\pi i} \delta_L {g_L}
D_L(\omega),\label{tad} \eeq where $\delta_L {g_L}$  can be found
by variation of Eq. (\ref{g_L}) in the field of
the $L$-phonon: \bea  \delta_L {g_L}&=&\delta_L {\cal F}
A(\omega_L){ g_L} + {\cal F} \delta_
L A(\omega_L){ g_L} \nonumber \\
&+& {\cal F} A(\omega_L)\delta_L{g_L}. \label{dgL} \eea

All the low-lying phonons we deal with  are of surface nature,  the surface peak dominating in their
creation amplitude: \beq g_L(r)=\alpha_L \frac {dU} {dr} +\chi_L(r). \label{gLonr}\eeq The first term
in this expression  is surface peaked, whereas the in-volume addendum $\chi_L(r)$ is rather small. It
is illustrated in Fig. 2 for the $2^+_1$ and $3^-_1$ states in $^{204}$Pb. If one neglects this
in-volume term  $\chi_L$, very simple expression for the non-pole term can be obtained \cite{scTFFS}:
\beq \delta\Sigma^{\rm non-pole}_L = \frac {\alpha_L ^2} 2 \frac {2L+1} 3 \triangle U(r).
\label{tad-L}\eeq Just as in \cite{levels,lev-semi}, below we will neglect the in-volume term in
(\ref{gLonr}) and use Eq. (\ref{tad-L}) for the non-pole term of $\delta \Sigma_L^{\rm PC}$.

All the low-lying phonons we consider have  natural parity. In
this case, the vertex $g_L$ possesses  even $T$-parity. It is a
sum of two components with spins $S=0$ and $S=1$, respectively, \beq
g_L= g_{L0}(r) T_{LL0}({\bf n,\alpha}) +  g_{L1}(r)
T_{LL1}({\bf n,\alpha}), \label{gLS01} \eeq where $T_{JLS}$ stand
for the usual spin-angular tensor operators \cite{BM1}. The
operators $T_{LL0}$ and $T_{LL1}$ have  opposite $T$-parities, hence
the spin component should be the odd function of the excitation
energy, $g_{L1}\propto \omega_L$. In all the cases we consider the component $g_{L1}$ is negligible
and the component $g_{L0}$ only will be retained, the index '0' being for brevity omitted: $g_{L0}\to g_L$.

As it was discussed in the introduction, any semi-magic nucleus consists of two sub-systems with
absolutely different properties, the normal and superfluid ones.
Following to \cite{lev-semi}, we consider the SP spectra only of the normal subsystem, e.g. the proton spectra
for the lead isotopes. In this case, all the above formulas remain valid with one exception.
Now the vertex $g_L(\bf r)$ obeys the QRPA-like TFFS equation \cite{AB}, \beq {\hat g}_L(\omega)=
 {\hat {\cal F}} {\hat A}(\omega) {\hat g}_L(\omega), \label{gL} \eeq
where all the terms are  matrices. The angular momentum projection $M$, which is written down in Eq.
(\ref{dSig-pole}) explicitly, is here and below for brevity omitted. In the standard TFFS notation, we
have: \beq {\hat g}_L=\left(\begin{array}{c}g_L^{(0)}
\\g_L^{(1)}\\g_L^{(2)}\end{array}\right),
\label{gs} \eeq \beq {\hat {\cal F}}=\left(\begin{array}{ccc}
{\cal F} &{\cal F}^{\omega \xi}&{\cal F}^{\omega \xi}\\
{\cal F}^{\xi \omega }&{\cal F}^\xi  &{\cal F}^{\xi \omega }\\
{\cal F}^{\xi \omega }&{\cal F}^{\xi \omega }& {\cal F}^\xi \end{array}\right). \label{Fs} \eeq

In (\ref{gs}), $g^{(0)}$ is the normal component of the vertex ${\hat g}$, whereas   $g^{(1),(2)}$
are two anomalous ones. In Eq. (\ref{Fs}), $\cal{ F}$ is the usual LM  interaction amplitude
which is the second variation derivative of the  EDF ${\cal E} [\rho,\nu]$ over the normal density
$\rho$ . The effective pairing interaction $\cal{ F}^\xi$ is the second derivative of the EDF over the
anomalous density $\nu$. At last, the amplitude ${\cal F}^{\xi \omega }$ stands for the mixed
derivative of ${\cal E}$ over $\rho$ and $\nu$.
\begin{figure}

\centerline {\includegraphics [width=80mm]{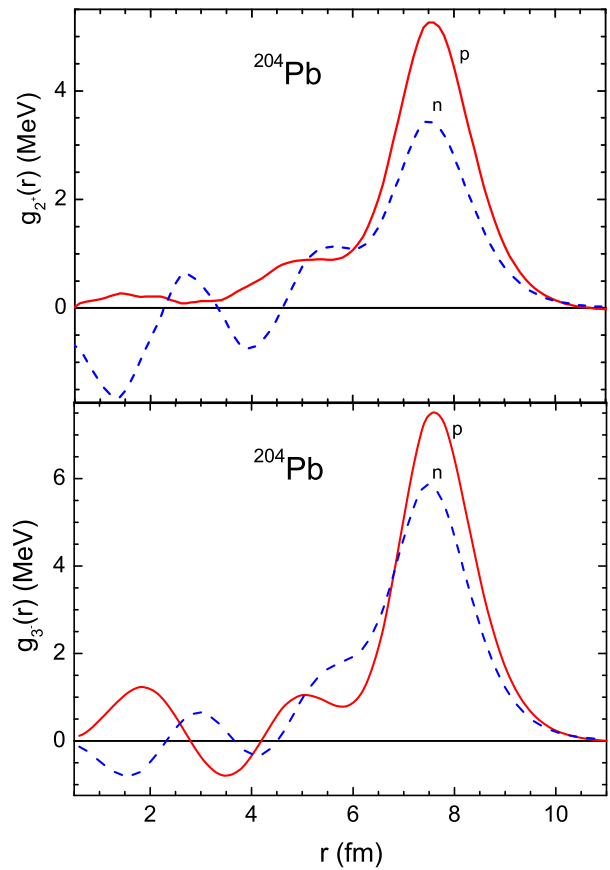}} \caption{(Color online) Phonon creation
amplitudes $g_L(r)$ for two low-lying phonons in the $^{204}$Pb nucleus.} \label{fig:gL}
\end{figure}

The matrix ${\hat A}$ consists of $3\times 3$  integrals over $\eps$ of the products of different
combinations of the Green function $G(\eps)$ and two Gor'kov functions $F^{(1)}(\eps)$ and
$F^{(2)}(\eps)$  \cite{AB}.

As we need the proton vertex ${\hat g}_L^p$ and the proton subsystem  is normal, only the normal
vertex $g_L^{(0)p}$ is non-zero in this case.  This is the meaning of the abbreviated notation $g_L$ in
(\ref{dSig-pole}) and below.

For solving the above equations, we use the self-consistent basis generated by the version DF3-a
\cite{DF3-a} of the Fayans EDF DF3 \cite{Fay4,Fay}. The nuclear mean-field potential $U(r)$ is the first
derivative of  ${\cal E}$ in (\ref{E0}) over $\rho$.

In magic nuclei \cite{levels}, the perturbation theory in $\delta \Sigma^{\rm PC}$ with
respect to $H_0$ can be  used to solve this equation: \beq \eps_{\lambda}=\eps_{\lambda}^{(0)} +
Z_{\lambda}^{\rm PC} \delta \Sigma^{\rm PC}_{\lambda\lambda}(\eps_{\lambda}^{(0)})
,\label{eps-PC0}\eeq where \beq Z_{\lambda}^{\rm PC} =\left({1- \left(\frac {\partial} {\partial \eps}
 \delta \Sigma^{\rm PC}(\eps) \right)_{\eps=\eps_{\lambda}^{(0)}}}\right)^{-1}. \label{Z-fac}\eeq

Another situation is typical for semi-magic nuclei, where small denominators
$\eps^{(0)}_1-\eps^{(0)}_2 \pm \omega_L$ can appear in Eq. (\ref{dSig-pole}). We demonstrate in Table
I the poor applicability of the perturbation theory on the example of the $^{204}$Pb nucleus with the
use of two phonons, $2_1^+, \omega_2{=}0.88\;$MeV and $3_1^-, \omega_3{=}2.79\;$MeV. The value of
$\delta \eps_\lambda^{\rm PC}$ in the column 6 is found as a sum of the values in three previous
columns multiplied, in accordance with Eq. (\ref{eps-PC0}), by the total $Z$-factor in the last
column.

The most dangerous denominator is now $\eps^{(0)}(2d_{3/2}){-}\eps^{(0)}(3s_{1/2}) {+}
\omega_2{=}0.09\;$MeV. The state 2$d_{3/2}$ is an obvious victim of the use of a non-adequate
perturbation theory in this case. Indeed, the inequality $1{-}Z_\lambda^{\rm PC}<<1$ is a necessary
condition for applicability of the perturbation theory.  For the 2$d_{3/2}$ state, we have the value
of 0.95 for the left hand side of this inequality, which is almost equal to unit. In the table, there
are other examples of poor applicability of the perturbation theory, but they are not so
demonstrative.

\begin{table*}
\caption{PC corrected proton SP energies $\eps_\lambda$ and $Z$-factors
of $^{204}$Pb nucleus found from the perturbation theory prescription (\ref{eps-PC0}) and (\ref{Z-fac}).}
\begin{tabular}{l c c c c c c c c c}
\hline
\hline\noalign{\smallskip}
$\lambda$ & $\eps^{(0)}_\lambda$ & $\delta \Sigma_{\lambda\lambda}^{\rm pole}$ &
$\delta \Sigma_{\lambda\lambda}^{\rm non-pole}$ & 'ghost' &   $\delta \eps_\lambda^{\rm PC}$ &
$\eps_\lambda^{\rm PC}$ & $Z_\lambda^{\rm PC}(L{=}2)$ & $Z_\lambda^{\rm PC}(L{=}3)$ & $Z_\lambda^{\rm PC}$ \\
\hline\noalign{\smallskip}
  1$i_{13/2}$  &   -1.21& -1.01  &  0.32 &  0.04 &-0.37  &-1.58 &0.64  &0.82  & 0.56      \\
  2$f_{7/2}$   &   -2.01& -0.81  &  0.20 &  0.03 &-0.36  &-2.37 &0.66  &0.90  & 0.62   \\
  1$h_{9/2}$   &   -3.36& -0.43  &  0.27 &  0.04 &-0.09  &-3.45 &0.74  &0.98  & 0.73 \\
  3$s_{1/2}$   &   -6.72&  0.07  &  0.17 &  0.03 & 0.22  &-6.50 &0.84  &0.96  & 0.82 \\
  2$d_{3/2}$   &   -7.51&  2.02  & 0.17 &   0.03 & 0.10 &-7.40  &0.05  &0.97  & 0.05  \\
  1$h_{11/2}$  &   -7.98&  0.28  &  0.30&   0.04 & 0.42 &-7.56  &0.70  &0.96  & 0.68  \\
  2$d_{5/2}$   &   -8.80&  0.33  & 0.18 &   0.03 & 0.26 &-8.541 &0.57  &0.80  & 0.50 \\
  1$g_{7/2}$   &  -10.93& -0.23  & 0.24 &   0.03 & 0.01 &-10.92 &0.79  &0.35  & 0.32  \\

\hline
\hline
\end{tabular}\label{tab1}
\end{table*}

For such cases, a method based on the direct solution of the Dyson equation (\ref{sp-eq}) was
developed in  \cite{lev-semi}. We describe it with the example of the same nucleus $^{204}$Pb which
was considered above to demonstrate the non-applicability of the perturbation theory. The model was
suggested based on the fact that the PC corrections are important only for the SP levels nearby the
Fermi surface. A model space $S_0$ was considered including two shells close to it, i.e. one hole and
one particle shells. To avoid any misunderstanding, we stress that this restriction concerns only Eq.
(\ref{sp-eq}). In Eq. (\ref{dSig-pole}) for $\delta\Sigma^{\rm pole}$, we use rather wide SP space
with energies $\eps_{\lambda}^{(0)}{<}40\;$MeV. The space $S_0$ involves 5 hole states (1g$_{7/2}$,
2d$_{5/2}$, 1h$_{11/2}$, 2d$_{3/2}$, 3s$_{1/2}$) and four particle ones (1g$_{9/2}$, 2f$_{7/2}$,
1i$_{13/2}$, 2f$_{5/2}$). We see that there is here only one state for each $(l,j)$ value. Therefore,
we need  only diagonal elements $\delta\Sigma^{\rm pole}_{\lambda \lambda}$ in (\ref{dSig-pole}),
which  simplifies very much  the solution of the Dyson equation.   As a consequence, Eq. (\ref{sp-eq})
reduces as follows: \beq \eps-\eps_{\lambda}^{(0)}-\delta \Sigma^{\rm PC}_{\lambda \lambda}(\eps) =0.
\label{sp-eq1}\eeq

The non-pole term does not depend on the energy, therefore only poles of Eq. (\ref{dSig-pole}) are the
singular points of Eq. (\ref{sp-eq1}). They can be readily found from (\ref{dSig-pole}) in terms of
$\eps_{\lambda}^{(0)}$ and $\omega_L$. It can be easily seen that the l.h.s of Eq. (\ref{sp-eq1}) always
changes sign between any couple of neighboring poles, and the corresponding solution
$\eps_{\lambda}^i$ can be found with usual methods. In this notation, $\lambda$ is just the index for
the initial single-particle state from which the state $|\lambda,i\rangle$ originated. The latter is a
mixture of a single-particle state with several particle-phonon states. The corresponding
single-particle strength distribution factors ($S$-factors) can be found similar to (\ref{Z-fac}):
\beq S_{\lambda}^i =\left(1- \left(\frac {\partial} {\partial \eps}
 \delta \Sigma^{\rm PC}(\eps) \right)_{\eps=\eps_{\lambda}^{i}}\right)^{-1}. \label{S-fac}\eeq
 Evidently, they should obey the normalization rule: \beq \sum_i S_{\lambda}^i =1.   \label{norm}\eeq
The accuracy of fulfillment of this relation is a measure of validity of the model space $S_0$ we use to
solve the problem under consideration.

\begin{table}[]
\caption{ Examples of solutions of Eq. (\ref{sp-eq1}) for protons in $^{204}$Pb.}
\begin{tabular}{ c  c  c   l }

\hline
\hline\noalign{\smallskip}
$\lambda$   & $i$ & $\eps_{\lambda}^i$  &\hspace*{7mm} $S_{\lambda}^i$   \\
\hline\noalign{\smallskip}
1$h_{9/2}$ &  1  &   -13.736  &  0.220 $\times 10^{-2}$ \\
           &  2  &   -11.596  &  0.777 $\times 10^{-3}$ \\
           &  3  &   -10.339  &  0.674 $\times 10^{-2}$ \\
           &  4  &    -8.862  &  0.484 $\times 10^{-3}$ \\
           &  5  &    -3.447  &  0.760 \\
           &  6  &    -2.217  &  0.199 \\
           &  7  &    -1.122  &  0.288 $\times 10^{-2}$ \\
\noalign{\smallskip}\hline\noalign{\smallskip}
            &     &   &$\sum S_{\lambda}^i{=}0.987$  \\
\noalign{\smallskip}\hline\noalign{\smallskip}
3$s_{1/2}$  &  1  &    -9.877  &  0.608 $\times 10^{-1}$ \\
            &  2  &    -8.536  &  0.604 $\times 10^{-1}$ \\
            &  3  &    -6.493  &  0.839 \\
\noalign{\smallskip}\hline\noalign{\smallskip}
            &     &   &$\sum S_{\lambda}^i{=}0.990$  \\
\noalign{\smallskip}\hline\noalign{\smallskip}
1$h_{11/2}$ &  1 &    -13.717  &  0.692 $\times 10^{-3}$ \\
            &  2 &    -11.690  &  0.256 $\times 10^{-1}$ \\
            &  3 &     -9.084  &  0.134 \\
            &  4 &     -7.509  &  0.814 \\
            &  5 &     -2.471  &  0.427 $\times 10^{-3}$ \\
            &  6 &     -1.095  &  0.473 $\times 10^{-2}$ \\
\noalign{\smallskip}\hline\noalign{\smallskip}
            &     &   &$\sum S_{\lambda}^i{=} 0.985$  \\
\noalign{\smallskip}\hline\noalign{\smallskip}
 2$d_{5/2}$ &  1  &   -11.817 &  0.314 $\times 10^{-2}$ \\
           &  2  &   -11.150 &  0.139 \\
           &  3  &    -9.799 &  0.516 $\times 10^{-1}$ \\
           &  4  &    -8.580 &  0.312 \\
           &  5  &    -8.195 &  0.295 \\
           &  6  &    -7.404 &  0.171 \\
           &  7  &    -0.564 &  0.791 $\times 10^{-3}$ \\
\noalign{\smallskip}\hline\noalign{\smallskip}
            &     &   &$\sum S_{\lambda}^i{=} 0.978$ \\
\noalign{\smallskip}\hline
 \hline

\end{tabular}\label{tab2}
\end{table}

\begin{figure}
\centerline {\includegraphics [width=80mm]{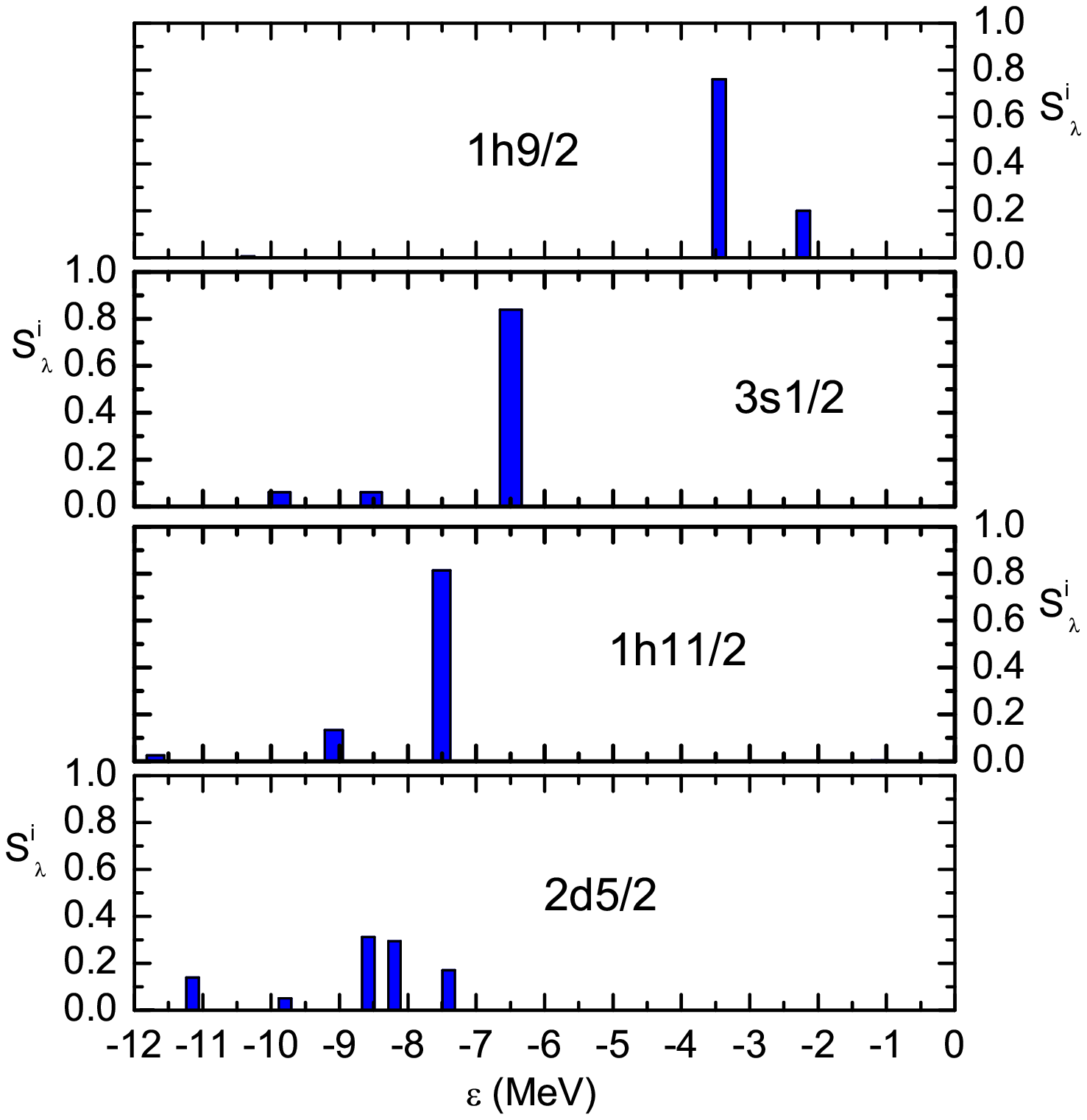}} \caption{(Color online) The SP strengh distributions
($S$-factors) of four  states in the $^{204}$Pb nucleus.} \label{fig:Slam}
\end{figure}

A set of solutions for four $|\lambda,i\rangle$ states in the $^{204}$Pb nucleus  is presented in
Table II. The corresponding $S$-factors are displayed in Fig. 3. In three upper cases for a given
$\lambda$ there is a state $|\lambda,i_0\rangle$ with dominating $S_{\lambda}^{i_0}$ value
(${\simeq}0.8$). They are examples of ``good'' single-particle states. In such cases, the following
prescription looks natural for the PC corrected single-particle characteristics: \beq
\eps_{\lambda}={\eps}_{\lambda}^{i_0}; \; Z_{\lambda}^{\rm PC}=S_{\lambda}^{i_0}. \label{i0} \eeq
These are the analogous to Eqs. (\ref{eps-PC0}) and (\ref{Z-fac}) in the perturbative  solution.

The lowest panel in Fig. 3 represents a case of a strong spread where there are two or more numbers
$i$ with comparable values of the spectroscopic factors $S_{\lambda}^i$. For such cases,  the
following generalization of Eq. (\ref{i0}) was suggested in \cite{lev-semi}:
\beq  \eps_{\lambda}= \frac 1 {Z_{\lambda}^{\rm PC}}
\sum_i {\eps}_{\lambda}^i S_{\lambda}^i,\label{spread1} \eeq
where \beq Z_{\lambda}^{\rm PC}=\sum_i
S_{\lambda}^i. \label{spread2}\eeq

The average SP energies and the $Z$-factors found according to the above prescriptions are presented
in Table III.

\begin{table}
\caption{PC corrected proton single-particle characteristics $\eps_\lambda$ and $Z_\lambda$ of
$^{204}$Pb nucleus found with Eqs. (\ref{i0}) or (\ref{spread1}) and (\ref{spread2}). The total
correction to the SPE $\delta \eps_{\lambda}^{\rm PC}{=}\eps_{\lambda}{-}\eps_{\lambda}^{(0)}$ is
presented.}
\begin{tabular}{l c c c c}
\hline
\hline
$\lambda$ & $\eps^{(0)}_\lambda$ &   $\delta \eps_\lambda^{\rm
PC}$ &$\eps_\lambda$ & $Z_\lambda$ \\
\hline
  1$i_{13/2}$  &   -1.21       &   0.14   &   -1.07   &   0.97  \\ 
  2$f_{7/2}$   &   -2.01       &  -0.23   &   -2.24   &   0.87  \\ 
  1$h_{9/2}$   &   -3.36       &   0.17   &   -3.19   &   0.96  \\ 
  3$s_{1/2}$   &   -6.72       &   0.23   &   -6.49   &   0.84  \\ 
  2$d_{3/2}$   &   -7.51       &   0.05   &   -7.46   &   0.94  \\ 
  1$h_{11/2}$  &   -7.98       &   0.25   &   -7.73   &   0.95  \\ 
  2$d_{5/2}$   &   -8.80       &   0.17   &   -8.63   &   0.92  \\ 
   1$g_{7/2}$   &  -10.93      &   0.13   &  -10.80   &   0.95  \\ 
\hline
\hline
\end{tabular}\label{tab3}
\end{table}

Comparison of Table III with the perturbation theory Table I shows limited common features. In
particular, the state  2$d_{3/2}$ is now absolutely ``healed'' with a big value of the $Z$-factor and
a small shift of the PC corrected energy from the initial value.

\section{PC corrections to the odd-even mass differences of semi-magic nuclei}
As it was discussed in Introduction, in heavy semi-magic nuclei the experimental data on the SP
energies are practically absent, but there is another kind of observables,  the odd-even mass
differences, which has a direct relevance to the set of the solutions $\eps_{\lambda}^i$  of the Dyson
equation in the form (\ref{sp-eq1}). In terms of the (TFFS) \cite{AB} they are named ``chemical
potentials'': \beq \mu_+^n(Z,N)=-\left(B(Z,N+1)-B(Z,N)\right),\label{mun_pl}\eeq \beq
\mu_-^n(Z,N)=-\left(B(Z,N)-B(Z,N-1)\right).\label{mun_mi}\eeq \beq
\mu_+^p(Z,N)=-\left(B(Z+1,N)-B(Z,N)\right),\label{mup_pl}\eeq \beq
\mu_-^p(Z,N)=-\left(B(Z,N)-B(Z-1,N)\right),\label{mup_mi}\eeq where $B(Z,N)$ is the binding energy of
the corresponding nucleus. Evidently, they are equal to one nucleon separation energies $S_{n,p}$
\cite{BM1} taken with the opposite sign. For example, we have $\mu_-^n(Z,N){=}{-}S_n(Z,N)$  or
$\mu_+^n(Z,N){=}{-}S_n(Z,N+1)$.

\begin{table}[b]
\caption{Excitation energies  $\omega_L$ (MeV) and the coefficients $\alpha_L$ (fm) in Eq.
(\ref{gLonr}) of the $2^+_1$ and $3^-_1$ phonons in even Pb isotopes.}
\begin{tabular}{c  c  c  c  c  c  c }
\hline
\hline\noalign{\smallskip} A  & $\omega_2$ & $\omega_2^{\rm exp}$  & $\alpha_2$
& $\omega_3$ & $\omega_3^{\rm exp}$    & $\alpha_3$    \\
\noalign{\smallskip}\hline\noalign{\smallskip}

180  &   1.415 & 1.168(1) &0.31 &  2.008 & -- &0.35 \\
182  &   1.284 &0.888  &0.31 &  1.836 & -- &0.35 \\
184  &   1.231 & 0.702 &0.32 &  1.839 & -- &0.36 \\
186  &   1.133 & 0.662 &0.33 &  1.881 & -- &0.34 \\
188  &   1.028 & 0.724 &0.34 &  1.968 & -- &0.34 \\
190  &   0.930 & 0.774 &0.36 &  2.052 & -- &0.33 \\
192  &   0.849 &0.854  &0.35 &  2.160 & -- &0.32 \\
194  &   0.792 & 0.965 &0.35 &  2.272 & -- &0.32 \\
196  &   0.764 & 1.049 &0.35 &  2.390 & 2.471(?) &0.31 \\
198  &   0.762 & 1.064 &0.35 &  2.506 & -- &0.31 \\
200  &   0.789 & 1.027 &0.30 &  2.620 & --     &0.31 \\
202  &   0.823 & 0.961 &0.31 &  2.704 & 2.517 &0.31 \\
204  &   0.882 & 0.899 &0.22 &  2.785 & 2.621 &0.31 \\
206  &   0.945 & 0.803 &0.16 &  2.839 & 2.648 &0.32 \\
208  &   4.747 & 4.086 &0.33 &  2.684 &2.615  &0.09 \\
210  &   1.346 & 0.800 &0.07 & 2.183  & 1.870(10) &0.19 \\
     &         &  &     & 2.587  &2.828(10)  &0.17  \\
212  &   1.444 & 0.805 &0.17 & 1.788  & 1.820(10) &0.36  \\
214  &   1.125 & 0.835(1) &0.19 & 1.469  & -- &0.37  \\

\noalign{\smallskip}\hline
\hline
\end{tabular}\label{tab1}
\end{table}

Indeed, let us write down the Lehmann spectral expansion for the Green function $G(\eps;{\bf r_1},{\bf
r_2})$ in the $\lambda$-representation of the functions which diagonalize $G$ \cite{AB}: \beq
G_{\lambda}(\eps){=}\sum_{s} \frac {|(a^+_{\lambda})_{s0}|^2}{\eps{-}(E_s{-}E_0){+}i\gamma}
{+}\sum_{s} \frac {|(a_{\lambda})_{s0}|^2}{\eps{+}(E_s{-}E_0){-}i\gamma},\label{Lehmann}\eeq with
obvious notation. The isotopic index $\tau=(n,p)$ in (\ref{Lehmann}) is for brevity omitted. In both
the sums, the summation is carried out for the exact states $|s\rangle$ of nuclei with one added or
removed nucleon. Explicitly, if $|0\rangle$  is the ground state of the even-even ($Z,N$) nucleus, the
states $|s\rangle$ in the first sum correspond to the ($Z,N+1$) one for $\tau=n$ and ($Z+1,N$) for
$\tau=p$. Correspondingly, in the second sum they are ($Z,N-1$) for $\tau=n$ and (Z-1,$N$) for
$\tau=p$. If $|s\rangle$ is a ground state of the corresponding odd nucleus, the corresponding pole in
(\ref{Lehmann}) coincides with one the chemical potentials (\ref{mun_pl}) -- (\ref{mup_mi}). At the
mean field level, they can be attributed to the SP energies $\eps_{\lambda}$ with zero excitation
energy, whereas with account for the PC corrections they should coincide with the corresponding
energies $\eps_{\lambda}^i$.

\begin{table*}[]
\caption{Proton odd-even mass differences $\mu_+$ (MeV) for the even Pb isotopes. The mean
field predictions for the Fayans EDF DF3-a and those with the PC corrections are given.}
\begin{tabular}{c c c c c c}
\hline
\hline\noalign{\smallskip}
nucl.& $\lambda$ & DF3-a &  DF3-a + $2^+$  & DF3-a + $(2^+{+}3^-)$  & exp \cite{mass}\\
\hline\noalign{\smallskip}
$^{180}$Pb&   1$h_{9/2}$ &  3.513   &  3.185 &   3.321 &    ---     \\

$^{182}$Pb&   1$h_{9/2}$ &  2.942   &  2.564 &   2.695 &    --- \\

$^{184}$Pb&   $1h_{9/2}$ &  2.360   &  1.906 &   2.093 &    1.527(0.094) \\

$^{186}$Pb&  $1h_{9/2}$ & 1.767    &  1.293 &   1.441 &    1.010(0.021) \\

$^{188}$Pb&  $1h_{9/2}$ & 1.172    &  0.683 &   0.806 &    0.461(0.031) \\

$^{190}$Pb&   $1h_{9/2}$ & 0.577    &  0.027 &   0.141 &   -0.112(0.020) \\

$^{192}$Pb&   $1h_{9/2}$ & -0.017   & -0.528 &  -0.420 &   -0.596(0.022) \\

$^{194}$Pb&   $1h_{9/2}$ & -0.608   & -1.167 &  -1.058 &   -1.107(0.023) \\

$^{196}$Pb&   $1h_{9/2}$ & -1.193   & -1.760 &  -1.658 &   -1.615(0.023) \\

$^{198}$Pb&   $1h_{9/2}$ & -1.769   & -2.316 &  -2.212 &   -2.036(0.025) \\

$^{200}$Pb&   $1h_{9/2}$ & -2.327   & -2.757 &  -2.648 &   -2.453(0.026) \\

$^{202}$Pb&   $1h_{9/2}$ & -5.806   & -5.177 &  -5.317 &   -5.480(0.039) \\

$^{204}$Pb&   $1h_{9/2}$ & -3.356   & -3.567 &  -3.447 &   -3.244(0.006) \\

$^{206}$Pb&   $1h_{9/2}$ & -3.818   & -3.911 &  -3.771 &   -3.558(0.004) \\

$^{208}$Pb&   $1h_{9/2}$ & -4.232   & -4.064 &  -3.959 &   -3.799(0.003) \\

$^{210}$Pb&   $1h_{9/2}$ & -4.670   & -4.653 &  -4.566 &   -4.419(0.007) \\

$^{212}$Pb&   $1h_{9/2}$ & -5.111   & -5.152 &  -4.980 &   -4.972(0.007) \\

$^{214}$Pb&   $1h_{9/2}$ & -5.555   & -5.686 &  -5.523 &   -5.460(0.017) \\

\hline\noalign{\smallskip}
 &$\langle \delta \mu_+ \rangle_{\rm rms}$ & 0.465  & 0.261&  0.252& \\
\hline
\hline
\end{tabular}\label{tab5}
\end{table*}
In the sets of solutions for four $|\lambda,i\rangle$ states in $^{204}$Pb presented in Table II, they
originate from the first hole and the first particle states in the model space $S_0$. In this case,
our prescription for the odd-even mass differences, in accordance with the Lehmann expansion
(\ref{Lehmann}), is as follows: \beq \mu_+(^{204}{\rm Pb})=-3.447\,\, {\rm MeV}, \label{muplPb204}\eeq
\beq \mu_-(^{204}{\rm Pb})=-6.493\,\, {\rm MeV}. \label{mumiPb204}\eeq

\subsection{The lead chain}
In this subsection, we consider the chain of even lead isotopes, $^{180-214}$Pb, with account of the two
low-lying phonons, $2^+_1$ and $3^-_1$. Their excitation energies $\omega_L$ and the coefficients
$\alpha_L$ in Eq. (\ref{gLonr}) are presented in Table IV. Comparison with existing experimental data
\cite{exp-omega} is given. We present only 3 decimal signs of the latter to avoid a
cumbersomeness of the table. On the whole, the $\omega_L$ values agree with the data sufficiently
well. In more detail, for the interval of $^{194-200}$Pb, the theoretical excitation energies of the
$2^+$-states are visibly smaller than the experimental ones.

This is a signal of the fact that our calculations may  overestimate the collectivity of these states
and, correspondingly, the PC effect in these nuclei. The opposite situation is present for the
lightest Pb isotopes, $A{<}190$, where we, evidently, underestimate the PC effect. The $\alpha_L$
value defines the amplitude, directly in fm, of the surface $L$-vibration in the nucleus under
consideration. We see that in the most cases both the phonons we consider are strongly collective,
with $\alpha_L{\simeq}0.3$ fm. At small values of $\omega_L$, the vibration amplitude behaves as
$\alpha_L {\sim}1/\omega_L$ \cite{scTFFS,BM2}.   Both the PC corrections to the SP energy, pole and
non-pole, are proportional to $\alpha_L^2$.   The ghost state $1^-$ is also taken into account,
although the corresponding correction for nuclei under consideration is very small, see
\cite{lev-semi}, because it depends on the mass number as $1/A$, $A{=}N{+}Z$.

\begin{table*}
\caption{Proton odd-even mass differences $\mu_-$ (MeV) for the even Pb isotopes.}
\begin{tabular}{c c c c c c}
\hline
\hline\noalign{\smallskip}
nucl.& $\lambda$ & DF3-a &  DF3-a + $2^+$  & DF3-a + $(2^+{+}3^-)$  & exp \cite{mass}\\
\hline\noalign{\smallskip}
$^{180}$Pb  &  3s$_{1/2}$ & -1.119   & -0.571 &  -0.793 &   -0.938(0.054) \\

$^{182}$Pb&    3s$_{1/2}$ &  -1.610  & -1.023 &  -1.268 &   -1.316(0.021) \\

$^{184}$Pb&    3s$_{1/2}$ & -2.104   & -1.450 &  -1.727 &   -1.753(0.022) \\

$^{186}$Pb&   3s$_{1/2}$ & -2.592   & -1.906 &  -2.152 &   -2.213(0.032) \\

$^{188}$Pb&   3s$_{1/2}$ &-3.072    & -2.356 &  -2.561 &   -2.661(0.019) \\

$^{190}$Pb&   3s$_{1/2}$ & -3.543   & -2.750 &  -2.945 &   -3.103(0.023) \\

$^{192}$Pb&   3s$_{1/2}$ & -4.005   & -3.265 &  -3.440 &   -3.572(0.020) \\

$^{194}$Pb&   3s$_{1/2}$ & -4.461   & -3.673 &  -3.838 &   -4.019(0.024) \\

$^{196}$Pb&   3s$_{1/2}$ & -4.911   & -4.111 &  -4.268 &   -4.494(0.025) \\

$^{198}$Pb&   3s$_{1/2}$ & -5.358   & -4.569 &  -4.716 &   -4.999(0.031) \\

$^{200}$Pb&   3s$_{1/2}$ & -5.806   & -5.177 &  -5.317 &   -5.480(0.039) \\

$^{202}$Pb&   3s$_{1/2}$ & -6.258   & -5.612 &  -5.753 &   -6.050(0.018) \\

$^{204}$Pb&   3s$_{1/2}$ & -6.717   & -6.357 &  -6.493 &   -6.637(0.003) \\

$^{206}$Pb&   3s$_{1/2}$ & -7.179   & -6.976 &  -7.105 &   -7.254(0.003) \\

$^{208}$Pb&   3s$_{1/2}$ & -7.611   & -7.778 &  -7.633 &   -8.004(0.007) \\

$^{210}$Pb&  3s$_{1/2}$ & -8.030   & -7.971 &  -8.055 &   -8.379(0.010) \\

$^{212}$Pb&   3s$_{1/2}$ & -8.446   & -8.276 &  -8.481 &   -8.758(0.044) \\

$^{214}$Pb&   3s$_{1/2}$ & -8.857   & -8.620 &  -8.865 &   -9.254(0.029) \\
\hline\noalign{\smallskip}
 &$\langle \delta \mu_- \rangle_{\rm rms}$ & 0.344   & 0.370 &  0.220& \\
\hline
\hline
\end{tabular}\label{tab6}
\end{table*}

The $\mu_+$ values, similar to that of (\ref{muplPb204}) for all the chain under consideration, are
given in Table V. In the last line, the average deviation is given of the theoretical predictions from
existing experimental data: \beq \langle \delta \mu_+ \rangle_{\rm rms} = \sqrt{ \sum (\mu_+^{\rm
th}-\mu_+^{\rm exp})^2/N_{\rm exp}}, \label{rms}\eeq where $N_{\rm exp}{=}16$. For comparison, we
calculated the corresponding value for the ``champion'' Skyrme EDF HFB-17 \cite{HFB-17} using the
table \cite{HFB-site} of the nuclear binding energies. It is equal to $\langle \delta \mu_+
\rangle_{\rm rms}({\rm HFB-17}){=}0.486$ MeV, that it is a bit worse than the DF3-a value even without
PC corrections. It agrees with the original Fayans's idea \cite{Fay1,Fay} to develop an EDF without PC
corrections. However, account for the PC corrections due to two low-laying collective phonons makes
agreement with the data significantly better. In this case, the phonon $2^+$ plays the main role, and
additional improvement of the agreement due to the phonon $3^-$ is not essential.

Let us go to the $\mu_-$ values similar to that of (\ref{mumiPb204}). For all the lead chain, from
$^{180}$Pb till $^{214}$Pb, they are given in Table VI. Again the last line contains the average
deviation of the theoretical predictions from the experimental data found from the relation similar to
(\ref{rms}) with the change of $\mu_+\to \mu_-$, and  $N_{\rm exp}{=}18$. In this case, the average of
the HFB-17 EDF predictions $\langle \delta \mu_- \rangle_{\rm rms}$(HFB-17)=0.253 MeV is better than
that of the DF3-a one, but again the account of the PC corrections permits to surpass the accuracy of
HFB-17. A relative role of the two phonons under consideration is now different from that in the case
of $\mu_+$, Table V. In the case of $\mu_-$, addition of the corrections due to the $2^+_1$ phonon
alone spoils the agreement a bit, but inclusion of both the phonons makes the agreement perfect. This
confirms the statement made in the Introduction about a delicate status of the problem under
consideration.

In conclusion of this subsection, we present the overall average deviation of the theoretical
predictions from the experimental data, summed over both the $\mu_+$ and $\mu_-$ values, $N_{\rm
exp}{=}16{+}18{=34}$. In this case, on the mean-field level, the average accuracy of the predictions
for the mass differences of the Fayans EDF $\langle \delta \mu \rangle_{\rm rms}({\rm
DF3{-}a}){=}0.389$ MeV is worse than that (0.333 MeV) for the Skyrme EDF HFB-17, but only a bit.
Account for the PC corrections due to the low-laying phonons $2^+_1$ and $3^-_1$ makes the agreement
significantly better, $\langle \delta \mu \rangle_{\rm rms}({\rm DF3{-}a})^{\rm PC}{=}0.218$ MeV.

\subsection{The tin chain}
Let us consider now the Sn chain, from $^{106}$Sn till $^{132}$Sn. To begin with, we consider validity
of the perturbation theory recipe (\ref{eps-PC0}) for the nucleus $^{118}$Sn, which is in the middle
of the chain. Table VII is completely similar to Table I for the nucleus $^{204}$Pb.  As it is clear
from the discussion of Table I, the $Z^{\rm PC}$ factors contain the main information on the
applicability of the perturbation theory for solving Eq. (\ref{sp-eq}). Table VII contains all
necessary information for this nucleus. The first seven columns contains a detailed information on the
PC corrections on the SP energies, whereas the last three ones, on the $Z$-factors.  The values in
columns 8, 9 and 10 are found from Eq. (\ref{Z-fac}) by substituting there the values of $\delta
\Sigma^{\rm PC}_{L=2}$, $\delta \Sigma^{\rm PC}_{L=3}$ and their sum $\delta \Sigma^{\rm PC}$,
correspondingly, in accordance with (\ref{sum-L}).
\begin{table*}
\caption{PC corrected proton SP energies $\eps_\lambda$ and $Z$-factors
of $^{118}$Sn nucleus found from the perturbation theory prescription (\ref{eps-PC0}) and (\ref{Z-fac}).}
\begin{tabular}{l c c c c c c c c c}
\hline
\hline\noalign{\smallskip}
$\lambda$ & $\eps^{(0)}_\lambda$ & $\delta \Sigma_{\lambda\lambda}^{\rm pole}$ &
$\delta \Sigma_{\lambda\lambda}^{\rm non-pole}$ & 'ghost' &   $\delta \eps_\lambda^{\rm PC}$ &
$\eps_\lambda^{\rm PC}$ & $Z_\lambda^{\rm PC}(L{=}2)$ & $Z_\lambda^{\rm PC}(L{=}3)$ & $Z_\lambda^{\rm PC}$ \\
\hline\noalign{\smallskip}

  1$h_{11/2}$   & -1.86 &-1.99& 0.54  & 0.06   &-0.53  &-2.39 &    0.57& 0.54  &  0.38     \\
  3$s_{1/2}$    & -2.50 &0.28 & 0.34 &0.05 &0.13 &-2.38  &0.19& 0.91  &  0.18    \\
  2$d_{3/2}$    & -2.64 &-6.20& 0.36 &0.08 &-0.19 & -2.83 &0.03 & 0.92   &  0.03    \\
  1$g_{7/2}$    & -3.67 &-1.27 &0.47 &0.07 &-0.35 & -4.02 &0.50 & 0.90   &  0.47    \\
  2$d_{5/2}$    & -4.49 &-2.09 &0.37 & 0.05 &-0.90 &-5.39   &0.60 & 0.84   &  0.54    \\
  1$g_{9/2}$    & -9.86 &0.13 &0.51 & 0.05&0.43 &  -9.43  &0.63 & 0.97   &  0.62   \\
\hline
\hline
\end{tabular}\label{tab7}
\end{table*}

We see that, just as for the nucleus $^{204}$Pb, Table I, there is a case of a catastrophic behavior
with the $Z$-factor close to zero, and again this is the state 2$d_{3/2}$. From the above
consideration for the lead chain, it is clear that for finding the $\mu_+$ and $\mu_-$ values we are
interested mainly on the first particle and first hole states   2$d_{5/2}$ and 1$g_{9/2}$,
correspondingly. Six states in Table VII create the model space $S_0$ in this case. In contrast to the
lead case, where particle and hole states have mainly opposite parities, now all the members of $S_0$
are of positive parity, with the only exception of the state 1$h_{11/2}$ which is rather distant and
plays no important role in the PC corrections. This occurs because of a unique feature of the
1$g_{9/2}$ level which alone is, in fact, a separate shell being an ``intruder state'' from the upper
shell. Usually, an  intruder state occurs among the states of the previous shell of states with
opposite parity, the $1f-2p$ shell in this case. However, the latter is rather distant from the
1$g_{9/2}$ level, and the inclusion of the $1f-2p$ shell into $S_0$ does not practically changes the
PC effects we analyze. Such a peculiar position of the hole shell consisted from the single 1$g_{9/2}$
state leads to  another peculiarity of the PC corrections in this state. Usually, the pole correction
$\delta \Sigma_{\lambda\lambda}^{\rm pole}$, see other states in this table and the Table I for
$^{204}$Pb, is rather big and negative, whereas the non-pole one is always positive but smaller than
the pole correction in absolute value, so that the total PC correction remains to be negative. In this
case the sign of the pole term is positive! Why it happens can be easily seen from Eq.
(\ref{dSig-pole}). Indeed, all close states of the positive parity which are coupled with the $2^+$
phonon, are now on the other side of the Fermi level, the second term of this expression, which
explains the positive sign of $\delta \Sigma_{\lambda\lambda}^{\rm pole}$. As a consequence, the total
PC correction to this SP level becomes positive.
\begin{table}[b]
\caption{Proton odd-even mass differences $\mu_+$ (MeV) for the even Pb isotopes. The mean
field predictions for the Fayans EDF DF3-a and those with the PC corrections are given.}
\begin{tabular}{c c c c c}
\hline
\hline\noalign{\smallskip}
nucl.& $\lambda$ & DF3-a &  DF3-a + $2^+$      & exp \cite{mass}\\
\hline\noalign{\smallskip}
$^{106}$Sn&   2$d_{3/2}$ &  0.295 &  -0.172&     -0.58851(0.00924) \\
$^{108}$Sn&   2$d_{3/2}$ &  -0.527&  -1.110&     -1.47008(0.01065) \\
$^{110}$Sn&   2$d_{3/2}$ &  -1.338&  -1.901&     -2.28372(0.02263) \\
$^{112}$Sn&   2$d_{3/2}$ &  -2.142&  -2.838&     -3.05006(0.01777)        \\
$^{114}$Sn&   2$d_{3/2}$ &  -2.936&  -3.484&     -3.73504(0.017)       \\
$^{116}$Sn&   2$d_{3/2}$ &  -3.720&  -4.230&     -4.83153(0.00853)         \\
$^{118}$Sn&   2$d_{3/2}$ &  -4.490&  -4.969&     -5.1103(0.0082)        \\
$^{120}$Sn&   2$d_{3/2}$ &  -5.245&  -5.748&     -5.78895(0.00371)        \\
$^{122}$Sn&   2$d_{3/2}$ &  -5.982&  -6.474&     -6.57225(0.00452)        \\
$^{124}$Sn&   2$d_{3/2}$ &  -6.700&  -7.066&     -7.31099(0.00361)     \\
$^{126}$Sn&   2$d_{3/2}$ &  -7.400&  -7.747&     -7.97316(0.01557)     \\
$^{128}$Sn&   2$d_{3/2}$ &  -8.084&  -8.342&     -8.55629(0.03889)      \\
$^{130}$Sn&   2$d_{3/2}$ &  -8.963&  -9.067&     -9.13802(0.00424)      \\
$^{132}$Sn&   2$d_{3/2}$ &  -9.892&  -9.838&     -9.66757(0.00608)      \\
\hline\noalign{\smallskip}
 &$\langle \delta \mu_+ \rangle_{\rm rms}$ &  0.722  &    0.286&    \\
\hline
\hline
\end{tabular}\label{tab9}
\end{table}

Another conclusion, which can be made from the analysis of Table VII, is the main role of the $2^+_1$
phonon for PC corrections of the tin nuclei. Indeed, the influence of a $L$-phonon can be estimated
from the value of the difference of $1{-}Z_L^{\rm PC}$. We see, that for all the states except
1$h_{11/2}$, the role of the $3^-_1$ phonon is rather moderate. Therefore, we shall find   the
odd-even mass differences $\mu_+$ and $\mu_-$ for all the tin chain with account only of the $2^+_1$
phonon, sometimes  checking the role of the $3^-_1$ phonon.

Table VIII contains the characteristics of the $2^+_1$ phonon which are used in the above calculation
scheme. The excitation energies $\omega_2$ are in reasonable agreement with experimental data. As in
the lead chain, we begin from the $\mu_+$ values.

\begin{table}[b]
\caption{Proton odd-even mass differences $\mu_-$ (MeV) for the even Pb isotopes. The mean
field predictions for the Fayans EDF DF3-a and those with the PC corrections are given.}
\begin{tabular}{c c c c c}
\hline
\hline\noalign{\smallskip}
nucl.& $\lambda$ & DF3-a &  DF3-a + $2^+$      & exp \cite{mass}\\
\hline\noalign{\smallskip}
$^{106}$Sn&   1$g_{9/2}$ &  -4.786&  -4.264&     -5.00208(0.01534)       \\
$^{108}$Sn&   1$g_{9/2}$ &  -5.652&  -5.014&     -5.79467(0.01651)     \\
$^{110}$Sn&   1$g_{9/2}$ &  -6.537&  -5.918&     -6.64302(0.0178)        \\
$^{112}$Sn&   1$g_{9/2}$ &  -7.424&  -6.669&     -7.55406(0.00408)      \\
$^{114}$Sn&   1$g_{9/2}$ &  -8.288&  -7.682&     -8.48049(0.00182)    \\
$^{116}$Sn&   1$g_{9/2}$ &  -9.102&  -8.544&     -9.27862(1.0809E-4)       \\
$^{118}$Sn&   1$g_{9/2}$ &  -9.859&  -9.320&     -9.99878(0.00538)       \\
$^{120}$Sn&   1$g_{9/2}$ &  -10.585& -10.024&   -10.68806(0.00821)       \\
$^{122}$Sn&   1$g_{9/2}$ &  -11.298& -10.743&   -11.39433(0.02982)      \\
$^{124}$Sn&   1$g_{9/2}$ &  -12.007&  -11.578&  -12.09275(0.02084)        \\
$^{126}$Sn&   1$g_{9/2}$ &  -12.713&  -12.291&  -12.82742(0.03747)        \\
$^{128}$Sn&   1$g_{9/2}$ &  -13.420&  -13.083&  -13.75269(0.03885)       \\
$^{130}$Sn&   1$g_{9/2}$ &  -14.128&  -13.905&  -14.58394(0.00482)        \\
$^{132}$Sn&   1$g_{9/2}$ &  -14.842&  -14.696&  -15.8073(0.00557)     \\
\hline\noalign{\smallskip}
 &$\langle \delta \mu_- \rangle_{\rm rms}$ &0.324  &  0.740  &    \\
\hline
\hline
\end{tabular}\label{tab10}
\end{table}

In the last line, just as in Tables V and VI, the average deviation is shown of the theoretical
predictions from the experimental data, found from Eq. (\ref{rms}). As for the Pb chain, we compare
the average accuracy of our calculations with those of the Skyrme EDF HFB-17. In this case we have
$\langle \delta \mu_+ \rangle_{\rm rms}({\rm HFB-17}){=}0.346\;$ MeV, which is significantly better
than the mean field prediction of the Fayans EDF DF3-a. However, the account for the PC corrections
makes the agreement for the latter significantly better, even better than that of the HFB-17 EDF.
\begin{table}
\caption{Characteristics of the $2^+_1$ phonons in even Sn isotopes, $\omega_2$ (MeV) and
$\alpha_2$(fm) in Eq. (\ref{gLonr}).}
\begin{tabular} { c  c  c  c  }
\hline
\hline\noalign{\smallskip}
$A$  & $\omega_2^{\rm th}$   & $\omega_2^{\rm exp}$ &  $\alpha_2$   \\
\noalign{\smallskip}\hline\noalign{\smallskip}
106    & 1.316      & 1.207    & 0.33    \\
108    & 1.231      & 1.206    & 0.37   \\
110    & 1.162      & 1.212    & 0.35  \\
112    & 1.130      & 1.257    & 0.40    \\
114    & 1.156      & 1.300    & 0.34     \\
116    & 1.186      & 1.294    & 0.32       \\
118    & 1.217      & 1.230    & 0.32     \\
120    & 1.240      & 1.171    & 0.33        \\
122    & 1.290      & 1.141    & 0.33     \\
124    & 1.350      & 1.132    & 0.27     \\
126    & 1.405      & 1.141    & 0.27      \\
128    & 1.485      & 1.169    & 0.23      \\
130    & 1.610      &  -       & 0.18       \\
132    & 4.329      & 4.041    & 0.22       \\
\noalign{\smallskip}\hline
\hline
\end{tabular}\label{tab8}
\end{table}
Let us go to the $\mu_-$ values which correspond to removing of a particle from an even Sn nucleus,
i.e. to adding a hole to the state 1$g_{9/2}$, see Table X. The above discussion of Table VII
indicates that one could expect something wrong in this case. And a bad event does occur: the PC
corrections spoil the agreement with the data, and significantly. This is especially disturbing as in
this case the mean-field predictions of the DF3-a EDF are rather good, on average better than those of
the HFB-17 EDF. Indeed, in this case the corresponding value of $\langle \delta \mu_- \rangle_{\rm
rms}$(HFB-17)=0.484 is significantly worse than the DF3-a value.

Our failure in the case of 1$g_{9/2}$ holes confirms the above remarks that inclusion of the PC
corrections from the low-lying phonons to the mean field  observable values found for the EDFs which
contain the phenomenological parameters involving such contributions on average is rather delicate
procedure. In this case, there is a special reason why it occurs, as far as this intruder state alone
generates a shell separated with the Fermi level from the particle shell of states of the same parity.
Such a situation is essentially different from the ``normal'' one when two shells adjacent to the
Fermi level, the particle and hole ones, contain mainly the states of opposite parities.

\section{Conclusion}

In this article, a method, developed recently \cite{lev-semi} to find the PC corrections to the SP
energies of semi-magic nuclei is used for finding the odd-even mass difference between the even
members of a semi-magic chain and their odd-proton neighbors. In terms of TFFS \cite{AB}, these
differences are the chemical potentials $\mu_+$ or $\mu_-$ for the addition or removal mode,
correspondingly.  The method  of \cite{lev-semi} is based on the direct solution of the Dyson equation
with the PC corrected mass operator $\Sigma^{\rm PC}(\eps){=}\Sigma_0{+}\delta \Sigma^{\rm PC}(\eps)$.

The Fayans EDF DF3-a \cite{DF3-a} is used  for finding the initial $\mu_{\pm}$ values and for
calculating the corresponding PC corrections. This is a small modification of the original Fayans EDF
DF3 \cite{Fay4,Fay}, the spin-orbit and effective tensor parameters being changed only. It should be
stressed that, from the very beginning \cite{Fay1,Fay4,Fay}, the Fayans EDF was constructed in such a
way that it should take into account all the PC effects on average. And, indeed, the DF3-a EDF turned
out to be rather accurate in predicting the $\mu_{\pm}$ values on the mean field level. We compare our
results with predictions of the Skyrme EDF HFB-17 \cite{HFB-17} fitted to nuclear masses with highest
accuracy among the self-consistent calculations. We analyzed four sets of data, $\mu_+$ and $\mu_-$
values in the Pb and Sn chains.

\begin{table}[h]
\caption{Average deviation $\langle \delta \mu_{\pm} \rangle_{\rm rms}$  of different theoretical predictions from the experimental data \cite{mass} for Pb and Sn chains.}
\begin{tabular}{c c c c}
\hline
\hline\noalign{\smallskip}
mode, chain     & HFB-17 &  DF3-a & DF3-a + PC\\
\hline\noalign{\smallskip}
$\mu_+$, Pb     &0.486  &0.465  &0.252 \\
$\mu_-$, Pb     &0.253  &0.344  &0.220 \\
$\mu_+$, Sn     &0.346  &0.722  &0.286 \\
$\mu_-$, Sn     &0.484  &0.324  & 0.740\\

\hline
\hline
\end{tabular}\label{tab11}
\end{table}
Table XI contains the values of the average deviation of the theoretical predictions from the data for
two EDFs, HFB-17 and DF3-a at the mean-field level and for the DF3-a EDF with phonon corrections. We
see that in two cases, the removal mode in Pb and the addition  one in Sn, the EDF HFB-17 exceeds
DF3-a in accuracy, but PC corrections to DF3-a make the agreement better compared to the HFB-17 EDF.
In the two other cases, the addition mode in Pb and the  removal  one in Sn, the mean field
predictions of the EDF DF3-a turned out to be more accurate than those of HFB-17. For the addition
mode in Pb chain PC corrections make the accuracy of DF3-a even better, but for the one in Sn chain a
failure occurs: PC corrections spoil the agreement, and significantly.The reason is discussed in
detail in the previous Section, but, in short, it is connected with a peculiar feature of the
$1g_{9/2}$ state which alone represents a separate shell. We think that such cases of ``bad'' behavior
of the PC corrections will be seldom, but this case demonstrates, that it is not clear {\it a priory}
if the PC corrections will improve agreement with experiment or not.  This agrees qualitatively with
the analysis of \cite{Dobaczewski} where it was found that the possibility of improvement of the
mean-field description of the SP energies with account for the PC corrections depends significantly on
the kind of the EDF used in these calculations.

\acknowledgments We acknowledge for support the Russian Science Foundation, Grants Nos. 16-12-10155
and 16-12-10161. The work was also partly supported  by the RFBR Grant 16-02-00228-a. E.S. and S.P.
thank the INFN, Sezione di Catania, for hospitality. This work has been carried out using computing
resources of the federal collective usage center Complex for Simulation and Data Processing for
Mega-science Facilities at NRC "Kurchatov Institute", http://ckp.nrcki.ru/. EES thanks the Academic
Excellence Project of the NRNU MEPhI under contract by the Ministry of Education and Science of the
Russian Federation No. 02. A03.21.0005.

{}


\begin{thebibliography}{}

\bibitem{Litv-Ring} E. Litvinova and P. Ring, Phys. Rev. C {\bf 73}, 044328 (2006).

\bibitem{Bort} Li-Gang Cao, G. Col\`o, H. Sagawa, and P.F. Bortignon,
 Phys. Rev. C {\bf 89}, 044314 (2014).

\bibitem{Dobaczewski} D. Tarpanov,  J. Dobaczewski,  J. Toivanen, and B.G. Carlsson,
Phys. Rev. Lett. {\bf 113}, 252501 (2014).

\bibitem{Baldo-PC} M. Baldo, P.F. Bortignon, G. Col\'o, D. Rizzo, and L. Sciacchitano,
J. Phys. G: Nucl. Phys.  {\bf 42}, 085109 (2015).

\bibitem{levels} N.V. Gnezdilov, I.N. Borzov, E.E. Saperstein, and S.V. Tolokonnikov,
Phys. Rev. C {\bf 89}, 034304 (2014).

\bibitem{scTFFS} V. A. Khodel, E. E. Saperstein, Phys. Rep. {\bf 92}, 183 (1982).

\bibitem{Khod-76} V. A. Khodel', Sov. J. Nucl. Phys.  {\bf 24}, 282 (1976).

\bibitem{phon-tad} S. Kamerdzhiev and E. E. Saperstein, EPJA {\bf 37}, 333 (2008).

\bibitem{tad} S. Weinberg, Phys. Rev. Lett. {\bf 31}, 494 (1973).

\bibitem{lev-exp} H. Grawe, K. Langanke, and G. Martínez-Pinedo,
Rep. Prog. Phys. {\bf 70}, 1525 (2007).

\bibitem{lev-semi} E.E. Saperstein, M. Baldo, S.S. Pankratov, and S.V. Tolokonnikov, JETP Lett. {\bf 104}, 609 (2016).

\bibitem{Ring-Werner}  P. Ring and E. Werner, Nucl. Phys. A {\bf 211}, 198 (1973).

\bibitem{AB}  A.B. Migdal {\it Theory of finite Fermi systems and applications to
atomic nuclei} (Wiley, New York, 1967).


\bibitem{chim-pots} E.E. Saperstein, M. Baldo, S.S. Pankratov, and S.V. Tolokonnikov, JETP Lett., to be published, {\bf 106}, (2017).

\bibitem{Fay1}
A.V. Smirnov, S.V. Tolokonnikov, S.A. Fayans, Sov. J. Nucl. Phys. {\bf 48}, 995 (1988).

\bibitem{Fay4}  I. N. Borzov, S. A. Fayans, E. Kromer, and D. Zawischa,
Z. Phys. A {\bf 355},  117 (1996).

\bibitem{Fay}
S.A. Fayans, S.V. Tolokonnikov, E.L. Trykov, and D. Zawischa, Nucl. Phys.  A {\bf 676}, 49 (2000).

\bibitem{DF3-a} S.V. Tolokonnikov and E.E. Saperstein, Phys.
At. Nucl. {\bf 73}, 1684 (2010).


\bibitem{BM1} A. Bohr and B.R. Mottelson, {\it Nuclear Structure} (Benjamin,
New York, Amsterdam, 1969.), Vol. 1.


\bibitem{HFB-17} S. Goriely, N. Chamel, and J. M. Pearson, Phys. Rev. Lett.
{\bf 102}, 152503 (2009).

\bibitem{HFB-site} S. Goriely, http://www-astro.ulb.ac.be/bruslib/ \\nucdata/



\bibitem{BE2} S.V. Tolokonnikov, S. Kamerdzhiev, D. Voytenkov,
S. Krewald, and E.E. Saperstein, Phys. Rev. C {\bf 84}, 064324 (2011).

\bibitem{Q-EPJA} S.V. Tolokonnikov, S. Kamerdzhiev,
S. Krewald, E.E. Saperstein, and D. Voitenkov, Eur. Phys. J. A {\bf 48}, 70 (2012).


\bibitem{Baldo-15} M. Baldo, P.F. Bortignon, G. Col\'o, D. Rizzo, and L. Sciacchitano,
J. Phys. G: Nucl. Phys. {\bf 42}, 085109 (2015).

\bibitem{exp-omega} S. Raman, C.W. Nestor Jr., and P. Tikkanen, At. Data Nucl. Data
Tables {\bf 78}, 1 (2001).

\bibitem{BM2} A. Bohr and B.R. Mottelson, {\it Nuclear Structure} (Benjamin,
New York, 1974.), Vol. 2.

\bibitem{mass}   M. Wang, G. Audi, A.H. Wapstra, F.G. Kondev, M. MacCormick, X.
Xu, and B. Pfeiffer,  Chinese Physics C, {\bf 36}, 1603 (2012).



\end{thebibliography}
\end{document}